\documentstyle[prl,aps,twocolumn,epsf]{revtex}
\begin{document}
\twocolumn[\hsize\textwidth\columnwidth\hsize\csname@twocolumnfalse%
\endcsname
\title{Self-duality in quantum impurity problems}
\author{P. Fendley$^1$ and H. Saleur$^2$}
\address{$^1$ Physics Department, University of Virginia,
Charlottesville VA 22901}
\address{$^2$ Department of Physics, University of Southern
California,
Los Angeles CA 90089-0484}
\date{April 1998}
\maketitle
\begin{abstract}

We establish the existence of an exact non-perturbative self-duality
in a variety of quantum impurity problems, including the Luttinger
liquid or quantum wire with impurity. The former is realized in the
fractional quantum Hall effect, where the duality
interchanges electrons with Laughlin quasiparticles. We discuss the
mathematical structure underlying this property, which bears an
intriguing resemblance with the work of Seiberg and Witten on
supersymmetric non-abelian gauge theory.

\end{abstract}
\pacs{PACS numbers: ???}  ]  Duality is a rather ancient concept.
It was probably observed first in electromagnetism, where the theory
is
invariant under exchange of electricity with magnetism. For the
semi-classical theory to
remain local, the electric charge $e$ and the magnetic charge $m$
obey
the Dirac quantization condition $em=2\pi$, so a large  $e$ (which
plays the role of the coupling) can be traded for a small $m$. In
general,
 duality  maps
a theory with strong coupling to one with weak coupling, and is thus
a powerful tool for exploring
strongly-coupled regimes.

A theory is self-dual when there is an exact map from a theory at
strong coupling to the {\sl same} theory at weak coupling. The
best-known example  occurs in  the two-dimensional Ising model, where
the map is known as Kramers-Wannier duality, and provides
a quick way to obtain the critical temperature.

In this paper, we present a exact form of self-duality in $1+1$
dimensional quantum impurity problems. This self-duality is much more
powerful than in the Ising model, because there is no
intermediate-coupling critical point interfering with the analytic
continuation from weak to strong coupling. This duality had sometimes
been used in the past as ``approximate'', with no clear understanding
of its exact status. We show here that it results from integrability,
a powerful symmetry which restricts drastically the
form of the Lagrangian in the low energy limit. We make the duality
very explicit by obtaining simple integral expressions for transport
properties. These hint at a deep mathematical structure, somewhat
similar to the one arising in supersymmetric gauge theories with
Seiberg-Witten duality \cite{SW}.

We study a model of $N$ flavors of interacting gapless fermions moving
in one dimension. We denote the left-moving and right-moving modes by
$\psi_{Li}(x)$ and $\psi_{Ri}(x)$ respectively with $i=1\dots N$. We
consider an impurity located at $x=0$ which gives a backscattering
interaction $L_B = \lambda \psi^{\dagger}_{Li}(0) \psi_{Ri}(0)+cc$.
These models can be bosonized in terms of $N$ free bosons, so that the
only interaction between left- and right-moving modes is at $x=0$.
For $N$=1, this is the Luttinger liquid with impurity, which in its
bosonic formulation is usually called the boundary sine-Gordon
model. For $N$=2, this is a quantum wire; the two species correspond
to the two spins up and down of the electrons. We consider only the
case where the $N$ species are identical, and we allow for charge
interactions, parametrized by a coupling constant $g$ ($g=N$ at the
free point).  Since the excitations are gapless in the bulk, the only
scale in the problem is the coupling $\lambda$ from the impurity
interaction.  There are two critical points: the ``no-backscattering''
point where $\lambda=0$, and the ``strong-coupling'' fixed point where
$\lambda\to\infty$ and all the excitations are backscattered.

Let us first discuss the simplest case $N$=1, the Luttinger liquid
with impurity. As is well known, the interacting fermions of the
Luttinger model can be bosonized in terms of a single free
boson. Moreover, one can ``fold'' space in half around the impurity at
$x$=0 and restrict to the half-line $0<x<\infty$
\cite{FLSbig}. Without the impurity, the Lagrangian is
$L_0=(1/4\pi g)\int_0^\infty dx\ (\partial_\mu\phi)^2$ where
$g$ parameterizes the interactions of the original Luttinger
fermions.  The two critical points correspond to Neumann boundary
conditions $\partial_x\phi(0,t)=0$ and Dirichlet boundary conditions
$\phi(0,t)=const$. As is familiar from conformal field theory, the
massless boson can be decomposed into left- and right-moving pieces
$\phi = \phi_L+ \phi_R$. One can then define the ``dual'' boson
$\widetilde\phi = (\phi_L - \phi_R)/g$ and rewrite the Lagrangian as $
L_0=(g/4\pi)\int_0^\infty dx\ (\partial_\mu\widetilde\phi)^2$.
Dirichlet boundary conditions on $\phi$ correspond to Neumann boundary
conditions on $\widetilde\phi$, and vice versa. With no interactions
at the boundary, the theory is therefore invariant under the mapping
$g\to 1/g$, and the interchange of Dirichlet and Neumann boundary
conditions.

An interesting set of charge-carrying operators are written in terms
of this boson and its dual as $ \Psi_{e,m}(x,t)=\exp[ie\phi(x,t) +
im\widetilde\phi(x,t)]$.  The parameters $e$ and $m$ must be integers
to ensure locality. For Euclidean time this entire problem can be
reformulated as a classical two-dimensional Coulomb gas \cite{Nien},
where $\Psi_{e,0}$ creates an electrical charge $e$, and $\Psi_{0,m}$
creates a magnetic charge $m$. In this context, the condition that
$e$
and $m$ be integers is analogous to the Dirac quantization
condition. The boundary two-point function is $\langle
\Psi_{e,m}(0,0)\Psi_{-e,-m}(0,t) \rangle \sim t^{-2x_{e,m}} $ where
the boundary scaling dimension $x_{e,m}= \left(e\sqrt{g} +
m/\sqrt{g}\right)^2$, which has the electric-magnetic duality $g\to
1/g$.

Including the impurity interaction causes operators to be added to
the
Lagrangian. A relevant boundary operator has boundary dimension
$x_{e,m}<1$. Without loss of generality we can consider
$g<1$, because any $g>1$ model can be mapped to a $g<1$ model by
sending
$g\to 1/g$ and exchanging the two fixed points. This duality is {\sl
not} self-duality; it does not say anything about the $g<1$ models by
themselves but only that the $g>1$ models are equivalent. When $g<1$,
only operators with $m=0$ are relevant \cite{KF}. When $g\ge 1/4$,
only
$\Psi_{1,0}$ and its conjugate $\Psi_{-1,0}$ are relevant. Therefore,
we consider the backscattering Lagrangian
\begin{equation}
L_B= \lambda(\Psi_{1,0}+\Psi_{-1,0}) = 2\lambda\cos\phi(0,t).
\label{lbound}
\end{equation}
This interaction induces a flow from Neumann boundary conditions (UV)
to Dirichlet (IR) as $\lambda$ increases, or, equivalently, as energy
decreases. From this expression one finds for weak backscattering at
zero temperature, the current $I$ and voltage $V$ obey the scaling
law
$I-gV \propto V^{2g-1}$ \cite{KF}.

The self-duality appears in the action in the strong coupling
limit. At very large $\lambda$, the field $\phi(0,t)$ is pinned at
the
minima of the potential, so it is reasonable to study the vicinity of
the Dirichlet fixed point by considering the instantons interpolating
between neighboring minima \cite{oldduality}. After some elementary
computations, one obtains the leading (charged) irrelevant operator
\begin{equation}
L_B \approx \lambda_D (\Psi_{0,1}+\Psi_{0,-1}) =2\lambda_D
\cos\widetilde\phi(0,t).
\label{ldual}
\end{equation}
The subscript $D$ is for dual, because this form of the
backscattering
lagrangian is dual to the form (\ref{lbound}).  Since
$\cos\widetilde\phi(0,t)$ has dimension $1/g$, $\lambda_D\propto
\lambda^{-1/g}$, and at zero temperature the current and voltage obey
the scaling law $I\propto V^{2/g -1}$ in the strong-backscattering
limit \cite{KF}. Thus the strong and weak-backscattering exponents
are
related by interchanging $g\to 1/g$.

If (\ref{ldual}) were exact, the theory would be completely self-dual
under the interchange $\lambda\to \lambda_D$ and $g\to 1/g$. This has
been often assumed in the literature, and is of course very
appealing,
especially in the context of the fractional quantum Hall effect at
$\nu=1/3$. Here, the Luttinger fermions at $g=1/3$ are equivalent to
Laughlin quasiparticles, and the impurity corresponds to a point
contact allowing tunneling between edges of a fractional quantum Hall
device \cite{KFreview}. In the weak-backscattering limit, the
tunneling (\ref{lbound}) is caused by Laughlin quasiparticles. In the
strong backscattering limit the system is effectively two different
Hall devices, weakly coupled by (\ref{ldual}) so that only electrons
can tunnel. Thus the duality maps a description in terms of Laughlin
quasiparticles to one in terms of electrons.

However, there is no reason a priori why the duality should be exact:
the instanton expansion of \cite{oldduality} is not exact, and
multi-instanton processes can be expected to contribute, adding to
(\ref{ldual}) terms $\propto \cos
n\widetilde\phi(0,t)$.  In addition, neutral operators will also
contribute, for instance the density - density coupling
$(\partial\widetilde\phi)^2$ (which has dimension 2 and so for
$g<1/2$
is the leading irrelevant operator). The perturbative
computations near the strong coupling fixed point will involve all of
these terms, and thus have little to do with its weak-coupling dual
(\ref{lbound}).

It turns out however that Bethe ansatz computations
\cite{FLSbig}
established that there {\sl is} indeed an
exact duality for the current at zero temperature:
\begin{equation}
I(\lambda,g,V)=gV-gI\left(\lambda_D,{1\over
g},gV\right)\label{dualcurr}
\end{equation}
Some well-motivated and well-checked conjectures require that this
also hold at arbitrary $T$ \cite{BLZ}.  The self-duality relation
(\ref{dualcurr}) can be derived directly by imposing the strong
constraints of integrability on the IR action. The integrability of
the
boundary sine-Gordon model means that there are an infinite set of
conserved quantities in involution, even with the perturbation
(\ref{lbound}) of the UV fixed point.  These are generated by the
non-local operator $\cos\widetilde\phi$, and by an infinite set of
local operators $O_{2k+2}$ which are, roughly, powers of the stress
energy tensor, with dimensions $2k+2$, $k$ an integer. By considering
the Yang-Baxter equation and its various algebraic consequences in
detail \cite{BLZ,hubert}, or more simply by demanding integrability
of
the flow {\sl both} near the UV and near the IR fixed point, it can
be
shown that the complete IR lagrangian is given by (\ref{ldual}) plus
an infinite series of the type $\sum_{k=0}^\infty c_k
\lambda_D^{(2k+1)g/(1-g)}O_{2k+2}$ (this sort of statement of course
can only make sense within a particular regularization scheme, here
dimensional regularization \cite{hubert}). The remarkable fact is
that
only {\bf one} charged operator appears.  It is also crucial that the
neutral operators all commute: as a result, it can be shown, by using
the non equilibrium Keldysh formalism, that DC transport properties
are insensitive to anything but (\ref{ldual}), and thus {\sl are}
self-dual, at any temperature. Therefore (\ref{dualcurr}) must hold
to
all orders in perturbation theory.

The existence of the duality prompts questions about the analytic
structure of the whole problem: knowing a priori that
(\ref{dualcurr})
holds, could one find the exact formula for $I$ without using the
Bethe ansatz? We present here a formulation of the problem that
suggests it should be the case, and bears an intriguing resemblance
with \cite{SW} (for a previous attempt in a different
direction, see \cite{weiss}).
Introducing the dimensionless parameter $u \propto
V\lambda^{1/(g-1)}$
(the precise relation is given in \cite{FLSbig}), doing perturbation
theory in powers of $\lambda$ in (\ref{lbound}) gives
\begin{equation}
\frac{I}{gV}= 1 - \sum_{n=1}^\infty a_n(g) u^{2n(g-1)}
\label{UVexp}
\end{equation}
for large $u$ (small $\lambda$).
The analogous expansion in the
strong-backscattering regime (small $u$) must be
\begin{equation}
\frac{I}{gV}= \sum_{n=1}^\infty a_n(1/g) u^{2n(1/g-1)}
\label{IRexp}
\end{equation}
because the self-duality relation (\ref{dualcurr}) means that not
only
the powers  but also the coefficients in the expansions of the
current are dual under $g\to
1/g$. Considering the dimensionless current ${\cal
I}(g,u)\equiv I/gV$ as an analytic function of $g$ gives the
non-perturbative self-duality
\begin{equation}
{\cal I}(g,u) = 1 - {\cal I}(1/g,u).
\label{dual}
\end{equation}
This equation applies for all $g$, so does much more than just relate
a model at $g>1$ to one with $g<1$.

Given either one of the expansions (\ref{UVexp}) or (\ref{IRexp}), we
can now write an expression for the current which not only makes
proving
the self-duality simple, but which displays an intriguing connection
to Seiberg-Witten theory.
The Bethe ansatz gives the coefficients $a_n$ to be \cite{FLSbig}
\begin{equation}
a_n(g) = \frac{(-1)^{n+1}}{n!}
\frac{\sqrt{\pi}}{2}\frac{\Gamma(ng+1)}{\Gamma(3/2
+n(g-1))}
\label{asubn}
\end{equation}
With this $a_n$, the expansions (\ref{UVexp}) and (\ref{IRexp}) have
finite radius of convergence, so the self-duality must hold
non-perturbatively. Simple expressions for the magnetization can be
found by using the integral form of the $B$ function
$\Gamma(a)\Gamma(b)/\Gamma(a+b)$ to rewrite $a_n(g)$. Since the sum
(\ref{UVexp}) is absolutely convergent for large enough $u$, we can
interchange the order of summation and integration, and do the
sum. This yields our result
\begin{equation}
{\cal I}(g,u)=\frac{i}{4u}
\int_{{\cal C}_0} dx \frac{1}{\sqrt{x + x^g - u^2}}
\label{Ilutt}
\end{equation}
where the contour of integration ${\cal C}_0$ starts at the origin,
loops around the square-root branch point on the positive real axis,
and returns to the origin. There must be such a branch point because
the expression inside the square root is positive for large $x$ and
negative at $x=0$.  By using the fluctuation-dissipation theorem
$\langle (\Delta I)^2\rangle = ((gV)^2/(1-g))
\partial_{V}{\cal I}$ derived in \cite{FLSnoise}, one obtains an
integral expression for the DC shot noise as well.

It can easily be checked that the series (\ref{UVexp}) can be
recovered from (\ref{Ilutt}) by changing variables in the integral to
$t=x/u^2$, and expanding the resulting integrand in powers of
$u^{2g-2}$. However, the integral in (\ref{Ilutt}) converges for all
real $u$, not just in the large-$u$ regime where the sum
(\ref{UVexp})
does. Therefore it must also yield the expansion (\ref{IRexp}) in the
small-$u$ limit.  By defining the variable $t=x/u^{2/g}$ and
expanding
the resulting integrand in powers of $u^{2/g-2}$, one can indeed
verify this fact.  Moreover, one can prove the self-duality relation
(\ref{dual}) without the series expansions, because the integral
expression (\ref{Ilutt}) is analytic in $g$ and so can be utilized
for
any $g$:
\begin{eqnarray}
{\cal I}(1/g,u)&=&
\frac{i}{4u}
\int_{{\cal C}_0} dx \frac{1}{\sqrt{x + x^{1/g} - u^2}}
\nonumber\\
&=&
\frac{i}{4u}
\int_{{\cal C}_0} dx \frac{ g x^{g-1}}{\sqrt{x + x^{g} - u^2}}
\nonumber\\
&=& \frac{i}{2u} \int_{{\cal C}_0}
dx\frac{\partial}{\partial x}
\left[
 \sqrt{x + x^{g} - u^2}\right]
- {\cal I}(g,u)\nonumber\\
&=& 1 - {\cal I}(g,u)
\label{dualproof}
\end{eqnarray}
where to get from the first to second line we change variables $x\to
x^{g}$, and ${\cal C}_0$
changes accordingly.

Our expressions for the current bear a strong resemblance with the
representations obtained in \cite{SW} for quantities such as mass gap
in $N$=2 supersymmetric gauge theory.  There the parameter $u$ is an
order parameter related to the expectation value of the Higgs field.
This is perhaps not so shocking.  The idea of Seiberg-Witten theory
is
to exploit the fact that certain physical quantities are analytic in
some coupling, except at specific singularities. If one takes the
coupling around a singularity, physical quantities do not necessarily
return to their original values (for example if they have logarithmic
behavior near the singularity): there is a non trivial  Berry's
phase, or monodromy. If one knows all the singularities and
monodromies
of an
otherwise-analytic function, one can reconstruct the complete
function
for any value of the coupling: its typical form in the present
context
is given by $\int dx/y$, where $y^2$ is a rational function of $x$,
and the integral loops around some of the square-root branch points
of
the integrand. The coefficients in $y^2$ depend on the couplings of
the theory. Singularities occur at values of the couplings where two
roots of $y^2$ coincide, so that the integral logarithmically
diverges.  In the gauge theories discussed in \cite{SW} and in many
subsequent generalizations, one can use perturbation theory to find
the monodromies. Thus perturbative information {\it and} analyticity
leads to non-perturbative results.

In integrable models, analyticity in couplings has long been exploited
to do various computations \cite{Baxter}. In the present context, the
analyticity is in the dependence of the current on complex $u$ (that
is, voltage). Little seems to be known in general about this
question,
but the integral expression (\ref{Ilutt}) reveals a very simple
structure.
The current is singular at the values $u_0$ where both $x + x^g -
u_0^2=0$ and $1+ gx^{g-1}=0$, namely $u_0^2=(-g)^{g/(1-g)}(1-g)$.
When $u=u_0$, two of the roots of $x+x^g-u^2$ coalesce. The integral
(\ref{Ilutt}) diverges if the contour runs in between the two
coalescing roots; this never happens for positive real $u$. Thus we
know the exact location of {\sl all} of the Lee-Yang zeroes of the
current. All of them have the same absolute value $|u_0|$, and for
rational values of $g$, there are a finite number (for some reasons
why this is, see \cite{paul}). The value $|u_0|$ gives the radii of
convergence: the series (\ref{UVexp}) absolutely converges for
$|u|>|u_0|$, while (\ref{IRexp}) absolutely converges for
$|u|<|u_0|$.

There are also non-trivial monodromies in this problem that can be
extracted from the integral expressions. For example, when $g=1/2$,
the integral can be done to find the known result ${\cal I}=1/2 -
(1/2u)\arctan(2u)$. This has singularities at $u_0=\pm i/2$. Looping
$u$ around one of the singularities moves onto a different sheet of
the arctangent, so ${\cal I}\to {\cal I} \pm \pi/(2u)$ for $g=1/2$.
Unfortunately, for general $g$ the monodromies are more complicated
and at the present time we do not know how to obtain them
without already knowing the integral (\ref{Ilutt}).

The foregoing self-duality extends to the case of $N$ channels. The
case $N$=2, a quantum wire with impurity, was studied perturbatively
in \cite{KF}, and the exact current was found using the Bethe ansatz
in \cite{LSS}.  Following \cite{FLSbig,LSS}, it is straightforward to
find the large-$u$ series expansion of the current for general
$N$. Defining as before ${\cal I}_N={I/gV}$, $g$ the conductance
without impurity ($g=N$ for $N$ channels of free electrons), one
finds
$${\cal I}_N =
\sum_{n=0}^\infty (-1)^{n} \frac{\sqrt{\pi}}{2}
\frac{\Gamma(n(1+Nh)+1)\Gamma(nh+1)}{n!\Gamma(nNh+1)\Gamma(nh+3/2)}
u^{2nh}
$$
where $h={1\over g}-{1\over N}$. This can be written as the integral
\begin{equation}
{\cal I}_N(g,u)=\frac{i}{4u}
\int_{{\cal C}_0} dx \frac{(1+x^h)^{N-1}}{\sqrt{x(1 + x^h)^N - u^2}}
\label{IN}
\end{equation}
This normalized current obeys the self-duality relation (\ref{dual})
as with the $N$=1 case, here with the dual coupling ${N^2/ g}$.  It
would be quite interesting to see whether this result generalizes to
the models discussed in \cite{YK}.

Duality considerations extend beyond transport properties; for
instance, it is known from the exact solution that there was a sort
of
duality between the UV and IR in the Kondo model \cite{Wiegmann}.
This
follows the same pattern as before.  Consider the anisotropic Kondo
problem, which is defined by the boundary Lagrangian generalizing
(\ref{lbound}), $L_B=\lambda(\Psi_{1,0}S_j^-+\Psi_{-1,0}S_j^+)$,
where $S^\pm$ are $su(2)_q$ spins ($q = e^{i\pi g}$, $g$
parameterizing
the anisotropy) in the representation of spin $j$. As with the
tunneling problem, the integrability requires the IR Lagrangian to
be very simple: it contains only the conserved quantities $O_{2k+2}$
and a single non-trivial term like (\ref{ldual}):
$L_B \approx \lambda_D
(\Psi_{0,1}S_{j-1/2}^-+\Psi_{0,-1}S_{j-1/2}^+)$, where now the spins
are in the representation of spin $j-1/2$.  As a
result, the free energy at magnetic field $H$ satisfies a duality
relation similar to (\ref{dual})
$$
f(j,\lambda,g,H)=f\left(j-{1\over 2},\lambda_D,{1\over
g},{H\over g}\right) + \dots
$$
The left-hand side is the expansion in the UV, whereas the right-hand
side is the IR expansion; the extra terms on the right-hand side are
only in odd-integer powers of $H/\lambda^{1/g-1}$, and arise because
the free energy does couple to the charge-neutral operators in the IR
action. The free energies for all $S$ are also given by
integrals like (\ref{Ilutt}), which surround branch points in the
complex plane \cite{paul}.

There are some interesting distinctions between these
quantum-impurity
problems and the Seiberg-Witten theories. In the gauge theories,
couplings never seem to appear in the exponents of $x$. Here $g$
rational leads to a polynomial for $y^2$ as in the gauge theories,
but
$g$ irrational is a slightly new situation. However, this is quite
interesting because none of the $g\to 1/g$ dualities known in gauge
theory have a proof like that in (\ref{dualproof}). Moreover, the
Luttinger liquid with impurity is not supersymmetric; as argued above
the analyticity seems to follow from the integrability, not from
supersymmetry as in the gauge theories. This opens up a new realm of
possibilities in gauge theory, a few of which are discussed in
\cite{paul}.


In conclusion,  it is  likely that
duality is a key feature of integrable field theories in 1+1
dimensions.  For example,  a situation  similar to what we have
discussed should occur  in the massless
flows between minimal models of conformal field theory, where we
expect
 a duality between the $\phi_{13}$ perturbations in the
ultraviolet and $\phi_{31}$ perturbations in the infrared.

\end{document}